# Heat release by isolated mouse brain mitochondria detected with diamond thermometer

A.M. Romshin[1,*], A.A. Osypov[2,3], I.Yu. Popova[2], V.E. Zeeb[2], A.G. Sinogeykin[4], I.I. Vlasov[1]

[1]Prokhorov General Physics Institute of the Russian Academy of Sciences, Moscow, Russia 119991.

[2]Institute of Theoretical and Experimental Biophysics of the Russian Academy of Sciences, Pushchino, Moscow Region, Russia 142292.

[3]Institute of Higher Nervous Activity and Neurophysiology of the Russian Academy of Sciences, Moscow, Russia 117485.

[4]Wonder Technologies LLC, Skolkovo Innovation Center, Bolshoy blvd. 42, Moscow, Russia.

e-mail: alex_31r@mail.ru

# Abstract

A production of heat by mitochondria is critical for maintaining body temperature, regulating metabolic rate and preventing oxidative damage to mitochondria and cells. Up to now mitochondrion heat production was characterized only by methods based on fluorescent probes which are sensitive to environmental variations (viscosity, pH, ionic strength, quenching etc.). Herein, for the first time the heat release of isolated mitochondria was unambiguously measured by a diamond thermometer (DT) which is absolutely indifferent to external non-thermal parameters. We show that during total uncoupling of transmembrane potential by CCCP application the temperature near mitochondria rises by 4-22 °C above the ambient temperature, with an absolute maximum of 45 °C. Such a broad temperature response may be associated with the heterogeneity of the mitochondria themselves as well as their aggregations in the isolated suspension. It also revealed spontaneous temperature bursts prior to CCCP application that can reflect involvement of some mitochondria to ATP synthesis or membrane potential leaking to avoid reactive oxygen species hyperproduction. The used temperature sensor and the data obtained shed light to the thermodynamics on the subcellular level.

# Introduction

Experimental detection of heat production in the single living cell assigned to its different compartments, organelles, ionic pumps and ionic channels has long been a cutting edge challenge for modern experimental approaches of cell physiology research. Proved existence of sharp steady-state temperature gradients (tens of degrees °C) in submicrometer watery volumes nearby nanoscale heat source [1] put forward beneficial base for thermal signaling concept [2], supposing that living cell are capable of creating significant temperature gradients at nanoscale, engaging thus ultralocal thermodynamic events as variable driving force to govern cascades of morphological, biochemical and electrical intracellular processes. Therefore, experimental uncovering of signaling role of the nanoscale pinpointed 3D thermodynamic events bound to ATP dependent biochemical and pumping processes, as well as to thermodynamics around ionic channels has an ultimate priority. Being the ubiquitous and specialized factory for transforming the most powerful chemical energy supply of the aerobic respiration process to the universal cellular energy source in the form of macroergic phosphate bonds of ATP, mitochondria are a natural producers of heat (up to 60 percent) that is generated due to inefficient energy conversion [3]. Together with decoupling mechanism that is used to control the upper value of the proton potential to avoid excessive production of reactive oxygen species [4], this served as a preadaptation to the development of homeothermy, that maintain the organism homeostasis and intensifies metabolism to the unpresidential level. This is drawn to the extreme in the brown fat of neonatants and hibernating animals, where mitochondria seem to be solely deduced to the heat production [5].

Various cellular dysfunctions and pathological conditions lead to changes in the energy metabolism of mitochondria [6] and, as a result, to changes in their heat production [7]. There is already a study on the use in medicine of the ability of mitochondria to heat up. Recently, thermosensitive nanocarriers have been developed for targeted drug delivery to mitochondria [8]. These nanocarriers are able to enhance the accumulation of anticancer drugs in mitochondria at a certain endogenous mitochondrial temperature. This increases the selectivity for cancer cells and helps to reverse cancer drug resistance [8]. Despite these facts, there are still unresolved questions about what is the absolute temperature of mitochondria from different cells/tissues/organs/organisms, what is the range of these temperature fluctuations, and how these fluctuations depend on the functional state in normal and pathological conditions.

A comprehensive understanding of the mitochondria thermogenesis requires reliable and robust technique without external biological impact. Recently, a great variety of techniques and approaches aimed to detect the mitochondrial temperature including fluorescent nanogels [9, 10], polymers [11] and proteins [12]–[14], molecular dyes [15]–[18] and colloidal quantum dots [19] were elaborated. A number of them using probes located outside of mitochondria (extramitochondrial) reported increases in temperature of several degrees caused by enhanced mitochondrial activity, generally, due to uncoupling of the respiratory chain which boosts mitochondrial thermogenesis. Another approaches based on thermosensors located within mitochondria (intramitochondrial) were able to identify an increase in the organelle temperature of several degrees in the same (or similar) conditions, suggesting the mitochondrial radiators could be warmer than their surroundings. Although Chr´etien et al. [15] revealed a mitochondrial temperature of more than 10 °C above that of the culture medium in normal conditions (without any stimuli), their study remains so far the only evidence of such a high heat release by mitochondria.

Despite numerous experimental works supporting the hypothesis of "hot mitochondria" [20, 21], the physical and biological mechanisms for furnishing perceptible thermal bursts remains a matter of controversy. So, Baffou et.al. argued that according to Fourier's law hardly such a small organelle (typically 500 nm in diameter) could heat the intracellular surrounding more than 0.1 mK [22]. Filling the temperature gap would necessitate organelles to produce the heat power $P = 1$ µW. Apparently, the mitochondria don't have enough fuel to meet this requirement. Baffou et. al. claimed the existing fluorescent thermosensors do not directly measure the temperature, and, like any fluorescent probe, are prone to many artifacts caused by variation in environment (viscosity, pH, ionic strength, quenching). Therefore, the validity of presented results along with their careful interpretation should be questioned.

In this paper, we employ a diamond thermometer (DT) designed by us for unambiguous extramitochondrial temperature measurements of mouse brain mitochondria. The DT based on a single fluorescent diamond microparticle fixed at the tip of the glass capillary and pre-calibrated by temperature is absolutely insensitive to external non-thermal parameters allowing unambiguous temperature detection. By means of DT instrumentation we demonstrate that the mitochondria isolated from the mouse brain could be several tens of degrees warmer than the surrounding environment at room temperature both during natural metabolic processes as well as under membrane permeability agent application. The simple physical model is also provided to support the obtained results.

# Materials and methods

All animal protocols and experimental procedures were performed in accordance with the requirements of Directive 2010/63/EC of the European Union and Order of the Ministry of Health of Russia of June 19, 2003 No. 267, and were approved by the Ethics Committees for Animal Experimentation at the ITEB RAS. BALB/c mice (25–33 g), purchased from the vivarium of laboratory animals Stolbovaya (https://www.pitst.ru, Moscow region, Russia), were housed with food and water *ad libitum*. Animals were maintained with a 12-h light/dark cycle (lights on from 9 AM to 9 PM) in a temperature-controlled room (22°C ± 1°C).

### Mouse brain mitochondria isolation

Adult mice (n = 6) were deeply anesthetized with isoflurane and decapitated. The hemispheres were excised and immediately placed into an ice-cold isolation buffer containing either 125 mM KCl and 10 mM $KH_2PO_4$, pH 7.4 (phosphate medium) or 220 mM mannitol, 70 mM sucrose, 10 mM hepes, 1 mM EGTA and 0.5 % bovine serum albumin (sucrose medium). The cerebellum was removed, and the rest of the brain tissue was cut into small pieces and placed in a 7 ml homogenizer (Duran, Wheaton) with 4 ml of the buffer and homogenized manually for 1 min with 20 strokes of tight-fitting pestle. The brain homogenate was centrifuged at 4000 g for 10 min at 4 °C. The pellet was discarded and the supernatant was centrifuged at 12000 g for 15 min at 4°C. The resulting pellet was resuspended in 0.5 ml of the isolation medium and stored on ice during the experiment. Hereafter, we apply the absolute mice (M) numeration where sucrose medium was used for M1-4 mice and phosphate medium for M5-6 mice.

## Mitochondria functional assay

Mitochondrial functional assay was carried out in the phosphate medium at room temperature 23 °C.

The 20 μM fluorescent dye TMRM (Tetramethylrhodamine, Methyl Ester, Perchlorate, ThermoFisher, USA) was used as the mitochondrial membrane potential indicator (the emission peak at 582 nm); 2 mM pyruvate (Sigma Aldrich, USA) was used as a respiration substrate, 100 μM ADP (Sigma Aldrich, USA) was used to activate ATP production, 4 μM mitochondrial oxidative phosphorylation uncoupler CCCP (2-[2-(3-Chlorophenyl) hydrazinylyidene] propanedinitrile, ThermoFisher, USA) was used to render mitochondrial inner membrane permeable to protons.

## DT design

The design principle of the diamond thermometer (DT) device was exhaustively described in [1]. Its backbone is represented by glass microcapillary with the low-strain single diamond crystal fixed at the tip of capillary. The thermosensitivity of DT is provided with an ensemble of silicon-vacancy centers (SiV-centers) embedded in the diamond crystal during the CVD-synthesis (see Methods). The position of the maximum of the SiV-fluorescence zero-phonon line (ZPL) is susceptible to the temperature and allows recording the temperature of any microsystem, having been calibrated only once.

## DT calibration

The calibration of the developed DT was performed ahead of the experiments with mitochondria. The DT was placed in an air environment for a preliminary determination of the power density of optical excitation at a wavelength that did not lead to a shift of the zero-phonon SiV luminescence line, and, consequently, to heating. The measured power density was used to determine the temperature dependence of the spectral position of the zero-phonon line center in a Linkam TS1500 thermostat, stabilized at a predetermined level with an accuracy of ~1 °C. The temperature in the chamber was changed in increments of 10 °C. At each step, the SiV luminescence spectrum was recorded and the spectral position of the zero-phonon line maximum was determined according to the algorithm proposed in our previous work [1].

Experimental setup

The study of mitochondrial uncoupling thermal response was performed using commercially available confocal spectrometer LabRam HR800 (Horiba). The SiV fluorescence was excited with a 473-nm laser light (Laser Quantum) which was focused by the low-NA objective to one of the ends of the multimode optical fiber (Thorlabs) with transmission maximum at 740 nm. Another fiber end laid through the interior of the capillary and located near its tip guided the excitation photons directly to microdiamond. The fluorescence was collected with a long-focal-length air objective (Olympus x50, NA=0.55) and directed to the spectrometer. Probing the mitochondrial viability with TMRM dye was implemented with the same optics equipped with a 532 nm laser and bandpass filter (550-700 nm) in the registration path.

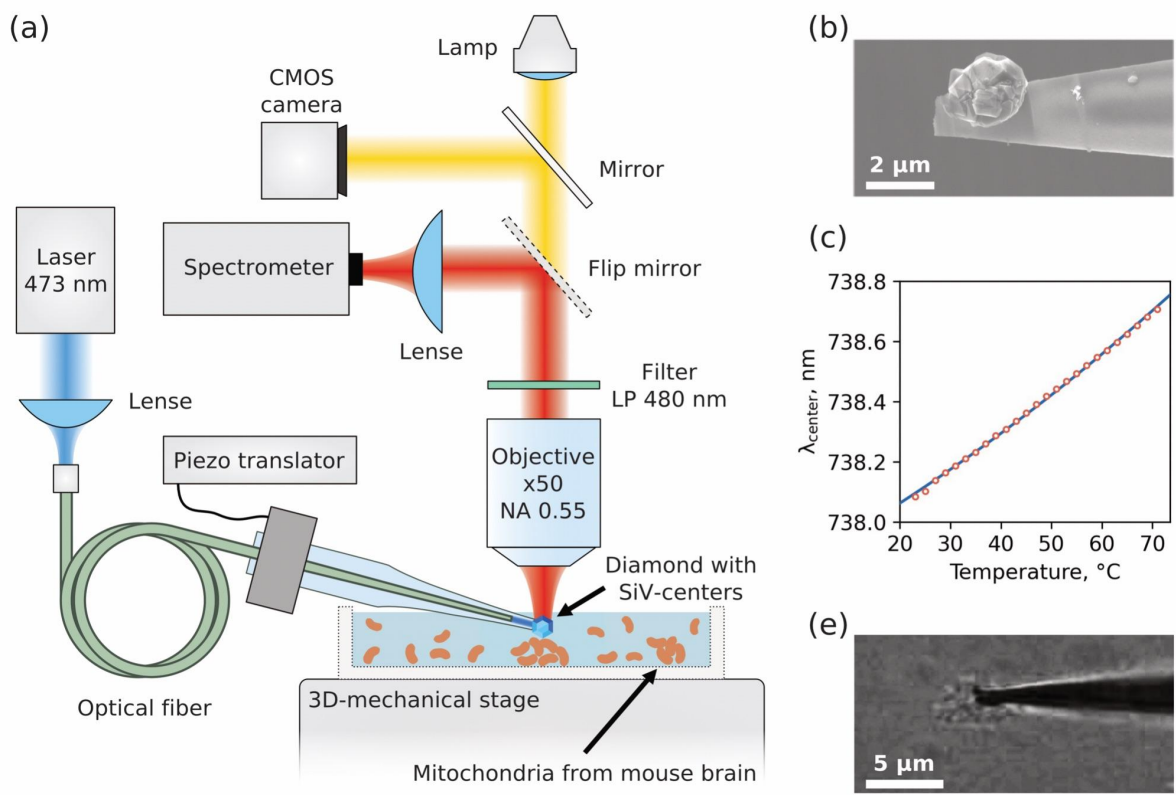

**Fig.1**. (a) Schematic drawing of the experimental setup employed for monitoring the SiV fluorescence and diffusion of mitochondria aggregates. The dish with mitochondrial suspension was coupled to the 3D-mechanical stage enabling suitable isolated aggregates of mitochondria to be easily found in CMOS-optical camera. Independently, the positioning of DT in the focal spot of the objective was governed with a three-axis micromanipulator; (b) SEM-image of the microdiamond evaluated at the capillary tip before melting; (c) Calibration dependence of the ZPL maximum position versus temperature in thermostat; (d) Representative optical image of the DT stuck to the pre-selected mitochondria aggregate.

# Results

The mitochondria used for studying the local heat production were isolated from 6 adult mice according to the protocol described in Methods. To verify the viability of the output organelles we examined electron transport chain with membrane potential sensitive dye. For this, immediately after isolation, mitochondria were loaded with TMRM dye which possesses bright and broadband fluorescence with maximum at 580 nm. The time-track of the dye intensity under application of pyruvate, ADP and CCCP was then recorded. Mitochondrial oxidative phosphorylation uncoupler CCCP renders mitochondrial inner membrane permeable to protons and thus spill off transmembrane potential with transformation of its energy directly to heat. Fig. 2 illustrates the typical bioenergetic profile for mitochondria isolated in sucrose (a) and phosphate (b) medium. At specified moments of time the organelles underwent substrat, ADP and CCCP application. The fluorescent changes $\Delta F/F_0$ were determined as normalized deviations from the baseline before any additions. While mitochondria isolated in both media demonstrate relevant physiological response reflected by changing of dye intensity with respect to membrane potential, the value of this response is higher in the sucrose medium, more favorable to mitochondrial physiological conditioning.

These measurements were performed for each mice once to verify that mitochondria survived after isolation and display conventional metabolic response.

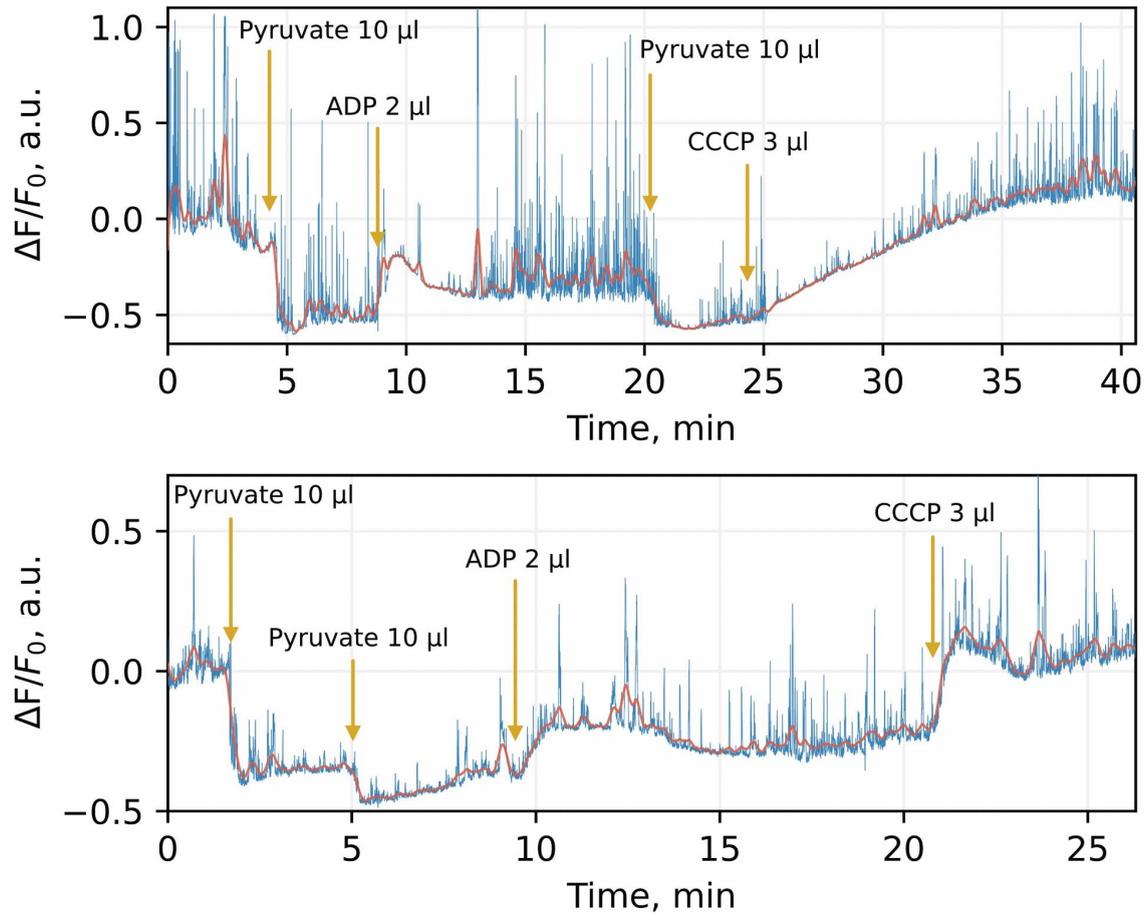

**Fig.2**. Changes of fluorescence 130 nM TMRM loaded with mitochondria isolated in sucrose-rich (a) and phosphate-rich (b) medium. The mixture of mitochondria suspension of 150 μl and 1 μl of TMRM was exposed to pyruvate (10 μl), ADP (2 μl) and CCCP (3 μl) action. The difference between the membrane potential reaction to respiration activation by substrate, its decrease during ATP production and the loss due to induced proton permeability by CCCP is clearly visible (note that the dye fluorescence decreases with the potential increase). The dye fluorescence was excited with a 532-nm laser and collected with the same optics as the SiV fluorescence.

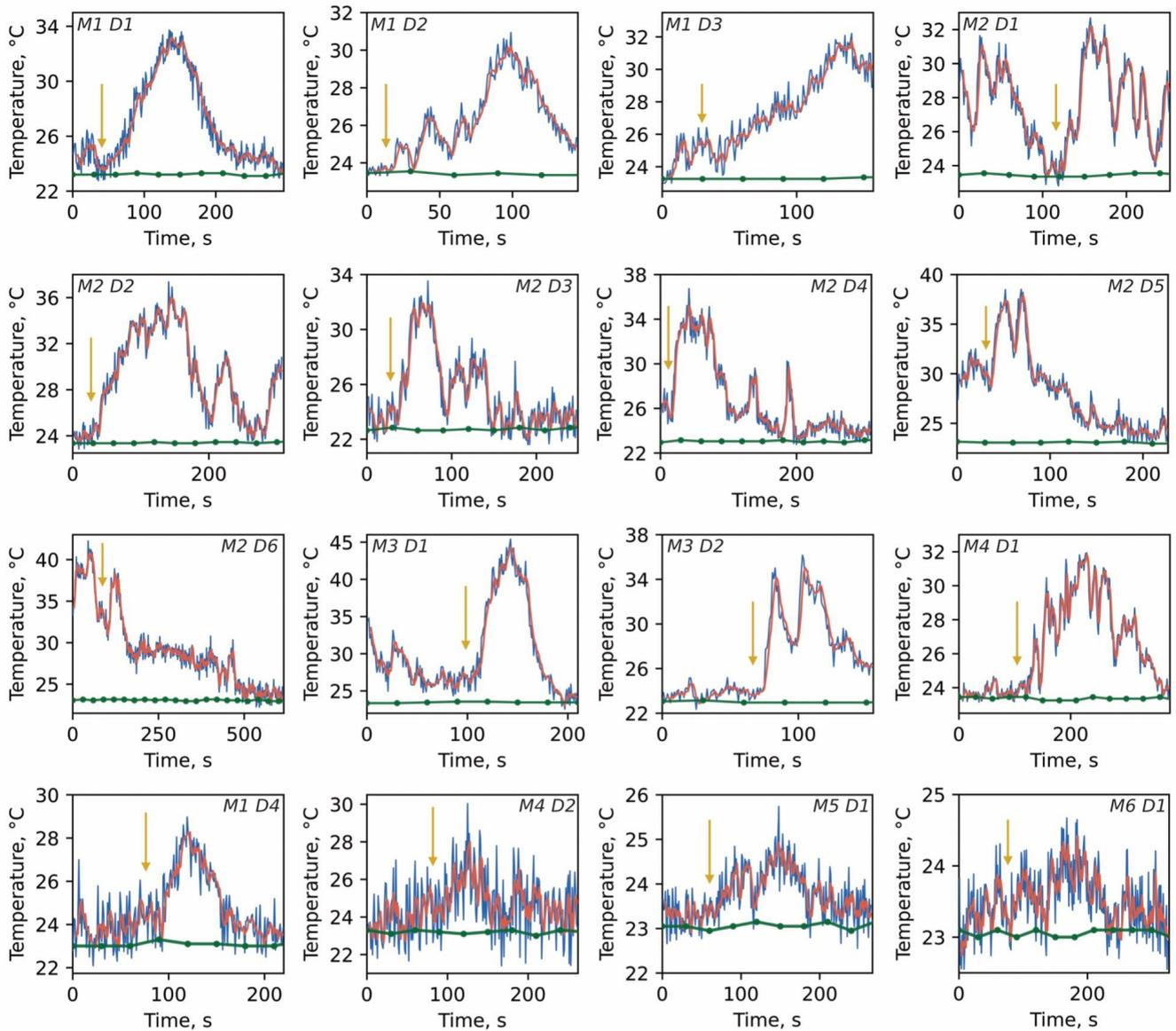

**Fig.3**. Time-tracks of the temperature recorded when DT was in close proximity of individual mitochondria aggregates. The primary data directly measured by DT is indicated in blue; smoothed data are given in red for eye guidance. The green curves are thermocouple temperature readings of the whole solution. The yellow arrows specify the moment of CCCP injection (2.5 µl). The text on top right-left corners specifies the mice (M) and dish (D) number for which the temperature was measured. Note the dish numbers are shown in chronological sequence with an average time between each one being ~20 min. The difference in the noise level is related to different thickness of the suspension layer and, therefore, dissimilar turbidity of the suspension.

Once the viability of organelles was confirmed the examination of mitochondria thermal response to the CCCP electro-chemical action started. 200 µl of mitochondrial suspension was applied to the cap of the Petry dish and smeared on its surface. The rest of the suspension was stored on ice throughout the experiment to preserve mitochondrial metabolism. The dish was then placed on the 3D-mechanical stage beneath the microscope objective and an appropriate aggregate of mitochondria with a size ranging from 2 to 10 μm was found through the optical CMOS-camera (see Fig. 1e; note that not all visible aggregates are mitochondria since some cell debris still remains after isolation). At the next step the DT was immersed in solution and the tip with microdiamond precisely touched the top surface of the targeted aggregate, slightly pressing on it. A little press is essential to leave the organelles immobilized during temperature protocols. To induce mitochondrial inner membrane permeability to protons and thus transmembrane potential spill off the 2.5 μl of CCCP was applied to the suspension and the temperature change was determined. As a reference indication of the macroscopic temperature we used a conventional thermocouple immersed in the solution. Before starting DT recordings, the temperature in the dish was determined as the steady-state ambient temperature.

Fig. 3 illustrates the temperature tracks recorded for different dishes within each of six mice. The time step between DT indications was chosen to be 1s for the reasons of the best signal-to-noise ratio and yet considering the possible quick thermal dynamics. One can see the temperature elevation $\Delta T$ ranging from several to tens degrees Celsius above the ambient level (green dots) primarily right after CCCP injection (yellow arrow). The maximum thermal burst within a single dish of $\Delta T_{max} = 22.4$ °C is observed for M3D1. However, temperature bursts were also detected before CCCP application (e.g., M2D1, M2D6, M3D1). Such spontaneous thermal responses are not lower in amplitude than CCCP-stimulated and could be attributed to the mitochondria which may be currently involved in ATP synthesis or controlled potential leaking to avoid reactive oxygen species hyperproduction due to excessive potential value.

The burst duration also varies in a wide range from seconds to hundreds of seconds (Fig. 4b), probably due to inhomogeneous distribution of CCCP action on individual mitochondria within one aggregate. While mitochondria on the edge of aggregate are exposed to the uncoupler action, the deep-dwelling organelles are isolated and remain unperturbed. Thus, the slow penetration of CCCP leads to asynchronous and heterogeneous time response.

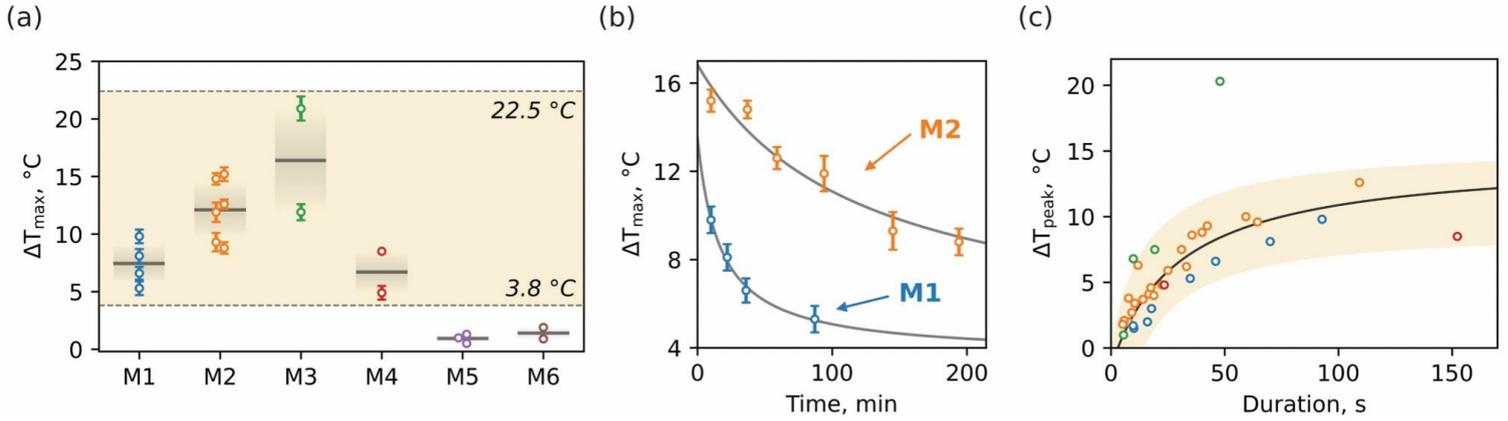

**Fig.4.** Statistics and kinetics of thermal responses per mice. (a) Maximal thermal response $\Delta T_{max}$ for a variety of mice. The gray lines indicate the mean value over dishes for selected mice, the same color gradients extend the standard deviation. The vertical dashed line separates two groups of mitochondria: M1-M4 - in glucose-containing and M5-M6 - in glucose free nutrient media. (b) Time evolution of $\Delta T_{max}$ for mice M1 and M2. Gray solid lines are added for eye guidance. (c) The amplitude of thermal bursts $\Delta T_{peak}$ collected from temperature traces as a function of their duration. Note, $\Delta T_{peak}$ takes into account the possible background humpback. Experimental points were fitted with black curve.

An analysis of temperature traces allowed us to extract $\Delta T_{max}$ per dish for a variety of mice (Fig. 4a). The mean thermal response of mitochondria for M1-M4 lies at the level of $<\Delta T_{max}> \sim 11$ °C though individual traces shows outstanding temperature breakthrough of 15-22 °C. In contrast, traces for M5 and M6 demonstrate less amplitude dynamics with mean $<\Delta T_{max}> \sim 1$ °C. The possible explanation of this behavior is based on the difference between mitochondria isolation media. The sucrose and mannitol containing medium with EGTA and BSA is more favorable to mitochondrial physiological conditioning during isolation period due to avoidance of excessive $K^+$ and phosphate load, $Ca^{2+}$ ions chelation, and free lipids absorbance which altogether leads to more yield of intact mitochondria with high metabolic capabilities.

Fig.4b demonstrates the alteration of $\Delta T_{max}$ in time for M1 and M2 with apparent decrease for both which is obviously due to mitochondria degradation over time passed. However, the attenuation rate for M1 is slightly faster than for M2. So, at 90 min thermal response amplitude reduces two times for M1 and only 1.3 times in case of M2. This dissimilarity may be determined by many factors including age and physiological condition of mice or slight differences during isolation especially at the homogenisation phase.

Ultimately, we delve into the kinetics of thermal responses for M1-M4 mice (Fig.4c). Each burst in temperature traces presented on Fig. 3 either spontaneous or CCCP-stimulated was carefully treated in order to evaluate temperature amplitude $\Delta T_{peak}$ in accordance with its duration determined as full-width at half-maximum (FWHM). The set of points calculated for individual mice are shown on Fig. 4c. The dependence clearly illustrates the saturation nature of thermal bursts and therefore was fitted (black curve) with $\Delta T_{peak}(\tau) = \frac{\Delta T_{peak}^{\infty} \tau}{\tau + \tau_{sat}}$, where $\Delta T_{peak}^{\infty}$ is temperature long-term limit, $\tau_{sat}$ is the saturation duration. An approximation gives $\Delta T_{peak}^{\infty} = 14.5$ °C and $\tau_{sat} = 33$ s purporting the principal amplitude of temperature bursts ceases to increase significantly at the duration of ~30 s. Apparently, this behavior implies that the resource of potential energy accumulated in mitochondria is limited and not enough to produce heat permanently.

# Discussion

### CCCP-induced heat production

Our experiments show that the transmembrane potential spill-out under CCCP application leads to mitochondria heating by 4-22 °C above the ambient temperature.

Such a broad temperature response may be associated with the heterogeneity of the mitochondria themselves as well as their aggregations in the isolated suspension. It is in accordance with the literature data of different mitochondrial types in neurons [23]. It was shown that small axonal mitochondria provide Ca ions bufferization to maintain sinapse transduction [24]. Dendrite mitochondria are rather elongated and the soma mitochondria form a kind of an organelle network [25]. There is data that mitochondria isolated from neuronal and glial cells differ by several physiological parameters [26]. Mitochondria from cells with greater energy demand have more cristae and less mitochondrial matrix volume. Therefore, the thermal profiles of mitochondria from different cellular and subcellular locations can be fundamentally different. Also, the spatial distance between DT active element and organelles aggregates changed randomly that could contribute to observed constant temperature dissimilarity, as the temperature gradient around them is rather steep.

The maximum absolute temperature that was recorded in our experiments is 45 °C. This value is close to 50 °C reported in the revolutionary work of Chre´tien et al. [15] and is the first independent confirmation of these results. Their study used two cell types: human embryonic kidney (HEK) 293 line, and primary skin fibroblasts. For the first time we report such high temperature values for brain mitochondria. Taken together, our and Chre´tien's data suggest the universal mitochondrial characteristics throughout all tissue types.

Rather intriguing is the temperature rise ceiling being apparently independent from the ambient temperature and cell type. In Chre´tien et al. [15] it was ~50 °C vs 38 °C and in our work it is ~45 °C vs 23 °C. In these cases the rise amplitude is different - is it due to various energy accumulation in the mitochondria in different experiments or rather due to a limited temperature endurance of the mitochondrial molecular machinery related to some kind of regulatory mechanisms keeping it in physiological boundaries? Do they yet vary depending on cell types or subcellular location? Further experiments are required to answer these questions.

The burst duration varies in a wide range from seconds to hundreds of seconds (Fig. 4b). This may be due to a well known heterogeneity of individual mitochondria in physical size, functional state, energy capacity, and other physiological parameters. Another source of the variation could be related to inhomogeneous CCCP action on individual mitochondria within aggregates. While mitochondria on the edge of aggregate are exposed to the uncoupler action, the deep-dwelling organelles are isolated and remain unperturbed. Thus, the slow penetration of CCCP leads to asynchronous and heterogeneous time response.

Taken together, this could explain the amplitude of thermal bursts as a function of their duration (Fig. 4c). In the low range below 50 sec it shows close to linear dependency with k ~0.3 °C/s and may be due to individual mitochondria variations, while at higher times the aggregate dynamics could come into play lengthening the response duration while the absolute temperature limit is already reached. The apparent outlier in M3 is probably the analysis artifact due to responses blended together and treated as a single one, although an exceptionally potent individual mitochondrion can not be ruled out.

Physiological heat production

Mitochondria are powerful and specialized factories for transforming the energy of the aerobic respiration process to the universal cellular energy equivalent in the form of macroergic phosphate bonds of ATP. Possessing a rather low coefficient of energy conversion efficiency in 40-60% they are natural producers of heat. They also maintain a decoupling mechanism that is used to control the upper value of the proton potential to avoid excessive production of reactive oxygen species. This mechanism renders the membrane permeable to protons and drops the potential with the releasing of heat due to the direct transformation of the electrochemical gradient energy to the thermal energy and serves as a preadaptation to the development of homeothermy, that maintain the organism homeostasis and intensifies metabolism to the unpresidential level. Thus in the normal physiological conditions we expect the increase in mitochondrial thermogenesis during ATP production and also in mitochondria with a good respiration substrates supply under the lack of ADP stock.

This shows up in our experiments as temperature bursts that were detected prior to CCCP application (e.g., M2D1, M2D6, M3D1). Such spontaneous thermal responses could be attributed to the mitochondria which may be currently involved in ATP synthesis or controlled potential leaking to avoid reactive oxygen species hyperproduction due to excessive potential value.

# Conclusions & Outlooks

The results presented shed light on the thermodynamics of mitochondria especially in mammalian central nervous system tissue. Mitochondrial thermodynamic studies are essential to understanding the fundamental mechanisms of neural tissue as these organelles play an important role in the control of neuroplasticity, including neuronal differentiation, neurite outgrowth, neurotransmitter release, dendritic and synaptic remodeling, and a host of other life processes. In addition to fundamental applications, a new methodological approach that allows temperature measurement at the subcellular level may finally reveal the unclear mechanisms underlying cerebral temperature dysregulation during and following neurovascular ischemia.

# Acknowledgments

This work was supported by RSF grant #20-65-46035 (to AO and IP).

# Declaration of interests

The authors declare no conflicts of interest.